\documentclass[twoside]{article}
\usepackage{qic}
\usepackage{amsmath}
\usepackage{pstricks}
\usepackage[]{youngtab}
\usepackage{bbm}

\renewcommand{\thefootnote}{\fnsymbol{footnote}}  

\newcommand{\F}{\ensuremath{\mathbbm{F}}}
\newcommand{\id}{\ensuremath{\mathbbm{1}}}
\newcommand{\ket}[1]{\mbox{$\vert#1\rangle$}}
\newcommand{\bra}[1]{\mbox{$\langle #1|$}}

\newcommand{\ketbra}[2]{\mbox{$\vert#1\rangle\!\langle #2\vert$}}
\newcommand{\entropy}[1]{\mbox{$H\!\left( #1\right)$}}
\newcommand{\bentropy}[1]{\mbox{$H_2\!\left( #1\right)$}}
\newcommand  {\BBP}{11.0028\%} 
\newcommand {\BBPn}{12.4120\%} 
\newcommand{\BBPub}{14.6447\%} 
\newcommand{\BBPcr}{12.9379\%} 
\newcommand{\BBPcq}{0.32656}   
\newcommand{\BBPcm}{500}       
\newcommand  {\SSP}{12.6193\%} 
\newcommand {\SSPd}{12.6904\%} 
\newcommand {\SSPn}{14.1119\%} 
\newcommand{\SSPcr}{14.5741\%} 
\newcommand{\SSPcq}{0.31210}   
\newcommand{\SSPcm}{250}       
\newcommand{\SSPer}{14.5930\%} 
\newcommand{\SSPeq}{0.31650}   
\newcommand{\SSPem}{300}       

\begin{document}
\setlength{\textheight}{8.0truein}    

\runninghead{Title  $\ldots$}
            {Author(s) $\ldots$}

\normalsize\textlineskip
\thispagestyle{empty}
\setcounter{page}{1}

\copyrightheading{0}{0}{2003}{000--000}

\vspace*{0.88truein}

\alphfootnote

\fpage{1}

\centerline{\bf
IMPROVED ONE-WAY RATES FOR BB84 AND 6-STATE PROTOCOLS}
\vspace*{0.37truein}
\centerline{\footnotesize
Oliver Kern and Joseph M.~Renes
}
\vspace*{0.015truein}
\centerline{\footnotesize\it Institut f\"ur Angewandte Physik,
Technische Universit\"at Darmstadt}
\baselineskip=10pt
\centerline{\footnotesize\it 64289~Darmstadt, Germany
}
\vspace*{0.225truein}
\publisher{(received date)}{(revised date)}

\vspace*{0.21truein}

\abstracts{
We study the advantages to be gained in quantum key distribution (QKD) protocols by combining the 
techniques of local randomization, or noisy preprocessing, and structured (nonrandom) block codes. Extending the results of [Smith, Renes, and Smolin, quant-ph/0607018] pertaining to BB84, we improve the best-known lower bound on the error rate for the 6-state protocol from 14.11\% for local randomization alone 
to at least 14.59\%. Additionally, 
we also study the effects of iterating the combined preprocessing scheme and find further improvements 
to the BB84 protocol already at small block lengths.
}{}{}

\vspace*{10pt}

\vspace*{3pt}
\communicate{to be filled by the Editorial}

\vspace*{1pt}\textlineskip    
\setcounter{footnote}{0}
\renewcommand{\thefootnote}{\alph{footnote}}
\section{Introduction}
Using a quantum channel to create a secret key between two parties is closely related to 
using the channel to send quantum information, with many
results found in one area applicable in the other. For instance, by treating the steps in a quantum
key distribution (QKD) protocol 
coherently and viewing the entire process as an entanglement distillation scheme, one can use 
properties of random quantum error-correcting codes to prove the security of the 
BB84 \cite{BB84} and 6-state \cite{Bru98} protocols up to 
bit error rates of $p_{max}=\BBP$ \cite{SP00} and $p_{max}=\SSP$  \cite{Lo01}, respectively.
Conversely, the formula for the quantum channel capacity can be obtained by importing the key rate 
resulting from a general approach to secret key generation over a known channel~\cite{devetak_relating_2004,devetak_private_2005,devetak_distillation_2005}.

One of the surprising results related to quantum capacity is the non-optimality of random codes, in
contrast to the classical case. The classical capacity of a channel can be achieved by using 
randomly-constructed block codes, and the independence of one input to the channel 
from the next results in a so-called single-letter formula for the capacity. 
Random coding can be used to create quantum error-correcting codes as well, but these do not always
achieve the capacity. Better performance
can be achieved by structured codes which exploit the ability of quantum error-correcting
codes to correct errors without precisely identifying them, a property called degeneracy. 
Initial results on rates achievable with degenerate codes displayed only modest gains \cite{ShSm96,DiSS98}, 
but recent analysis shows that degeneracy is crucial to the behavior of optimal codes \cite{SmSm07}.

By appealing to the coherent formulation of the protocol, degenerate codes should also be useful in QKD. 
This was shown to be the case in the original security proof of 
the 6-state protocol \cite{Lo01}, as the results of
\cite{DiSS98} were used to improve the error rate threshold to $p_{max}=\SSPd$. 
More striking threshold improvements are possible, if counterintuitive, by simply adding noise to 
the raw key bits before they are processed into the final key, a procedure known
as local randomization \cite{KGR05,KGR05b}. This 
improves the error rate thresholds for the two protocols to $p_{max}=\BBPn$  and 
$p_{max}=\SSPn$, respectively. At first glance, these results make no sense in the coherent 
picture of QKD, since adding more noise to already noisy entangled pairs only decreases the amount
of pure entanglement which can be extracted. 
The entanglement/secret-key analogy does not hold perfectly, however; entangled states are sufficient, but
not necessary, for creation of secret keys. A broader class of states, called private states, lead to 
secret keys when measured~\cite{horodecki_secure_2005}, and these should properly be the target output 
of the coherent version of the QKD protocol.  
Indeed, the exact error thresholds are recovered in the coherent picture when the 
QKD protocols with local randomization are analyzed in these terms \cite{RenesSmith07}. 

With a systematic understanding of how degenerate codes and local randomization boost the key rate, it 
becomes sensible to combine the two methods to look for even higher thresholds. 
Recently it was shown in \cite{SRS06} that doing so improves the error threshold of the BB84 protocol 
up to at least $p_{max}\approx 12.92\%$ by using the same type of structured code studied in~\cite{ShSm96,DiSS98,SmSm07}. These specific codes consist of the concatenation of two codes, 
the first a simple repetition
code and the second a random code. The repetition code, sometimes called a cat code in the context of 
quantum information theory since the codewords are $\ket{0}^{\otimes m}$ and $\ket{1}^{\otimes m}$, induces
degeneracy in the overall code since a phase flip on any of the physical qubits leads to the same logical error, and
is corrected in the same way. 
In particular, blocklength $m=400$ corresponds to the threshold stated above.
Since the random code portion of the protocol 
corresponds to information reconciliation and privacy amplification in the
classical view, the local randomization and the repetition code together become a
type of \emph{preprocessing} performed before these ``usual'' steps.  
In this paper we show that the same preprocessing protocol as used in \cite{SRS06} 
can also be used to improve the maximum tolerable bit error rate for the 6-state protocol, up
to at least $p_{max}=\SSPer$ for a blocksize of $m=\SSPem$.
This is already quite close to the upper bound of  $\BBPub$ \cite{FGGNP97,KGR05,moroder_one-way_2006} 
on the tolerable error rate
for the BB84 protocol, and since the error threshold grows with blocklength, the bound is
presumably exceeded at larger blocklengths, 
indicating the higher robustness of the 6-state protocol.
In addition we investigate iterating the preprocessing scheme in the BB84 protocol, 
and show an improvement both in rate and error threshold over single-round preprocessing 
for even modest blocklengths. Our calculations are facilitated by a closer look at
the representation theory relevant to describing the quantum states resulting from the preprocessing,
enabling us to continue the investigation started in~\cite{SRS06} to the 6-state protocol and 
iterated versions of the preprocessing for BB84. 

To begin, section II describes the preprocessing scheme in more depth and then derives 
secret key rate expressions for the BB84 and the 6-state protocols. Numerical 
calculations for blocklengths into the hundreds are then presented for the two protocols. 
Section III examines the advantages of iterating the preprocessing protocol to achieve higher
rates and thresholds for the same amount of effort in noise addition and block coding. The appendix
explains how representation theory is helpful for the numerical evaluation of such key rates in both cases.

\section{Secure key rates using the preprocessing protocol}
\begin{figure}[t]
 \centering
 \includegraphics[scale=0.8]{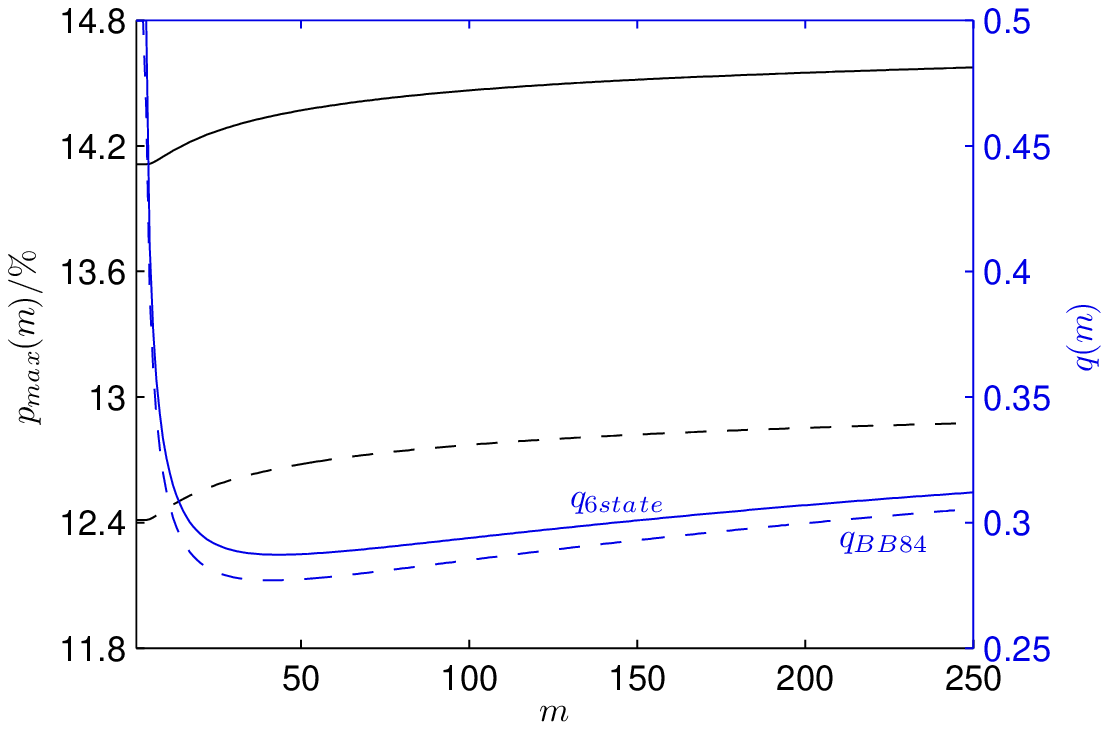}
 \vspace*{13pt}
 \fcaption{
Maximum tolerable bit error rate $p_{max}$ (left y-axis, black) and the corresponding rate $q$ of the added noise for which it is achieved (right y-axis, blue) versus block length $m$.
Dashed lines correspond to the BB84 protocol, solid lines to the 6-state protocol.
 \label{fig:1}}
\end{figure}

\noindent
The preprocessing protocol proposed in \cite{SRS06} begins after Bob has received the quantum signals
from Alice and they have sifted their raw keys to throw out mismatches between the preparation 
and measurement basis. 
Alice then flips each of her sifted key bits $(x_1,\dots,x_n)$
with probability $q$, resulting in new bits $(\tilde{x}_1,\dots,\tilde{x}_n)$. These
are partitioned into blocks of size $m$, and for each block she computes the syndrome 
$(\tilde{x}_1\oplus\tilde{x}_2,\tilde{x}_1\oplus\tilde{x}_3,\dots,\tilde{x}_1\oplus\tilde{x}_m)$ and 
sends this information to Bob. He 
computes the \emph{relative} syndrome of their blocks by adding his corresponding syndrome to Alice's,
modulo two. Alice's message is public knowledge, but the first bit of each block is still secret, 
so it is kept as a potential key bit. 
The protocol then proceeds with the usual error correction and privacy amplification 
steps to transform these kept bits into a secret key, 
now aided by the relative syndrome of each block and knowledge of the probability $q$ of
local randomization.   
Without local randomization, it turns out that $m=5$ is the optimal blocklength for
improving the error threshold in the 6-state protocol---longer blocklengths have worse thresholds~\cite{Lo01}. 
However, the results in \cite{SRS06} indicate that with the addition of noise, 
the highest tolerable bit error rate of BB84 grows with the blocksize $m$, and we 
find a similar result in the 6-state case (see figure \ref{fig:1}). 

We determine the secure key rates of the BB84 and 6-state one-way key distillation protocols involving the preprocessing protocol described above using the security proof of Renner \cite{phdrenner}.
This proof states that the key rate of such a protocol is given by
\begin{equation}\label{eq:rennerrate}
 r = \frac{1}{m} \min_{\sigma_{AB} \in \Gamma} \bigl( S(X|E) - S(X|Y) \bigr)
\end{equation}
where the minimum ranges over the set of states $\Gamma$ of all density operators on the $2\times 2$ dimensional Hilbert space $\mathcal{H}_A \otimes \mathcal{H}_B$ such that the measurement performed during the parameter estimation phase of the protocol leads to a certain bit error rate $p$. 
The conditional von Neumann entropies in \eqref{eq:rennerrate} are calculated for the states
\begin{equation}\label{eq:sgima_xye}
 \sigma_{XY\overline{E}} = \mathcal{E}_{XY\overline{E}\leftarrow A^mB^mE^m} (\sigma_{ABE}^{\otimes m} )
\end{equation}
which describe the processing of each block, including local randomization and syndrome calculation, 
and eventual measurement of the output qubits of the repetition code. That is, the preprocessing is treated
quantum-mechanically or coherently, but the usual processing classically. Here $X$ denotes
Alice's key outcome when measuring the output bits and $Y$ Bob's key and syndrome outcomes. 

For the BB84 protocol the set $\Gamma$ contains the states
\begin{equation}\label{eq:sigma_ab}
 \sigma_{AB} = \sum_{u,v} p_{uv} X^u_B Z^v_B \ketbra{\Phi^+}{\Phi^+} Z^v_B X^u_B,
\end{equation}
where $\ket{\Phi^+}_{AB}=\frac{1}{\sqrt{2}}\sum_k \ket{kk}_{AB}$ and 
$\{ p_{uv} \} \equiv \{p_{00},p_{10},p_{11},p_{01}\} = \{1-2p+t,p-t,t,p-t\}$, $t\in [0,p]$. In 
the 6-state protocol, meanwhile, parameter estimation assures us that $\Gamma$ contains only the single state $\sigma_{AB}$ with $\{ p_{uv} \} = \{ 1-\frac{3}{2}p, \frac{p}{2}, \frac{p}{2}, \frac{p}{2}\}$.

Using Renner's proof allows us to include the preprocessing but still only minimize over the quantum 
states $\sigma$ corresponding to individual signals. The crucial simplification is that the quantum state
of the block can be taken to be the product $\sigma^{\otimes m}$ without loss of generality. Other
proof techniques would require minimization over all possible 
(potentially-entangled) block states, or an additional
step in the parameter estimation procedure to ensure that the state does have this power form.

\subsection{Computation of the secure key rate}

To compute the secure key rates we make use of the fact that the difference of entropies in \eqref{eq:rennerrate} can also be written as difference of corresponding quantum mutual informations, i.\,e.~$S(X|E) - S(X|Y) = I(X:Y)-I(X:E)$.
In order to calculate these quantities, we need to determine the states $\sigma_{XY\overline{E}}$ defined in \eqref{eq:sgima_xye}.
Start with an $m$-fold tensor product of a purification of $\sigma_{AB}$,
\begin{equation}
\ket{\sigma}_{ABE} \equiv \ket{\sigma}^{\otimes m}_{ABE_1E_2} =
\sum_{\vec{u},\vec{v}} \sqrt{ p_{\vec{u},\vec{v}} }  X^{\vec{u}}_{B} Z^{\vec{v}}_{B},
\ket{\Phi^+}^{\otimes m}_{AB} \ket{\vec{u}}_{E_1}\ket{\vec{v}}_{E_2},
\end{equation}
where $\vec{u},\vec{v}\in\{0,1\}^m=\F^m$ so that $X^{\vec{u}}=X^{u_1}\otimes X^{u_2}\otimes\cdots$,
and similarly for $Z^{\vec{v}}$ and $p_{\vec{u},\vec{v}}$.
We now need to calculate the state resulting from noisy preprocessing followed by a blockwise stabilizer code measurement in which the stabilizers contain Pauli $\id$ and $Z$ operators only.

The first step, local randomization, can be described in a coherent way by adding 
a classical register $\mathbf{A'}$ (such systems will be denoted with boldface type) in the state
$( (1-q)\ketbra{0}{0}+q\ketbra{1}{1} )^{\otimes m}$ and then applying controlled not gates from 
the individual register states to the bits $A$. 
This leads to 
\begin{equation}\label{eq:sigmap}
\ket{\sigma'}_{ABE} =
 \sum_{\vec{u},\vec{v},\vec{f}} \sqrt{ p_{\vec{u},\vec{v}} q_{\vec{f}} }
 X^{\vec{u}+\vec{f}}_{B} Z^{\vec{v}}_{B}
 \ket{\Phi^+}^{\otimes m}_{AB} \ket{\vec{f}}_{\mathbf{A'}} \ket{\vec{u}}_{E_1} Z^{\vec{f}}_{E_2}\ket{\vec{v}}_{E_2},
\end{equation}
where $\vec{f}\in\F^m$ and $q_{\vec{f}}=q^f(1-q)^{m-f}$ for $f=|\vec{f}|$, the number of 1s in $\vec{f}$, a notation we shall use throughout. Here we have used the fact that $X_A\ket{\Phi^+}_{AB}=X_B\ket{\Phi^+}_{AB}$ to simplify the expression; this move is responsible for the $Z^{\vec{f}}$ operation applied to $E_2$.

In the second step, Alice and Bob both measure the $m-1$ (generators of the) stabilizers of a $\id/Z$-only stabilizer code 
which encodes one logical qubit into $m$ physical qubits.
Using a public (authenticated) channel, Alice sends her syndrome to Bob who calculates the relative syndrome $\vec{s}$ by adding Alice's string to his measurement outcome modulo two.
Afterwards both decode their encoded state.
Such a stabilizer code together with an encoding
$U_{\rm enc} \ket{ \vec{e},c } = \ket{\theta( \vec{e},c )}$,
where
\begin{equation}\label{eq:encodedstabstate}
 \ket{\theta(\vec{e},c)} = \ket{ c\cdot\vec{\eta}_m + \sum_{j=1}^{m-1} e_j\cdot \vec{\eta}_j },
\end{equation}
can be fully specified by defining two bases
$\{ \vec{\xi}_i \}_{i=1\dots m}$ and
$\{ \vec{\eta}_j \}_{j=1\dots m}$ both spanning $\F^m$ with the property that
$\vec{\xi}_i \cdot \vec{\eta}_j = \delta_{ij}$ (see e.\,g.~\cite{Ha03Fi}).
In this case the stabilizers are given by $Z(\vec{\xi}_i) = Z^{\vec{\xi}_i}$, $i=1\dots m-1$, and a measurement of these stabilizers on the encoded state \eqref{eq:encodedstabstate} will give the syndrome $\vec{e}$.
Measurement of the logical $Z$ operator $Z(\vec{\xi}_m)$ gives the value of the encoded bit $c$.
Applying one of the $X(\vec{\eta}_i) = X^{\vec{\eta}_i}$, $i=1\dots m-1$, operators on a encoded state results in a flip of the $i$-th bit of the syndrome, while applying the logical $X$ operator $X(\vec{\eta}_m)$ flips the encoded bit, $c\mapsto c\oplus 1$.
Both the set of all $Z(\vec{\xi}_i)$ and the set of all $X(\vec{\eta}_j)$ are complete sets of commuting observables.
Note that (i)
\begin{equation}
 \ket{\Phi^+}^{\otimes m}_{AB} = \frac{1}{\sqrt{2^{m-1}}} \sum_{\vec{e}} \frac{1}{\sqrt{2}} \sum_{c}
 \ket{\overline{\theta}(\vec{e},c)}_A \ket{\theta(\vec{e},c)}_B,
\end{equation}
and (ii) that any $m$ fold Pauli operator can be decomposed as
\begin{equation}
 X^{\vec{u}} Z^{\vec{v}} =
                   X^{\vec{\xi}_m \cdot \vec{u}}(\vec{\eta}_m) Z^{\vec{\eta}_m  \cdot \vec{v}}(\vec{\xi}_m)
\prod_{i=1}^{m-1}  X^{\vec{\xi}_i \cdot \vec{u}}(\vec{\eta}_i) Z^{\vec{\eta}_i  \cdot \vec{v}}(\vec{\xi}_i),
\end{equation}
where $\vec{\xi}_m \cdot \vec{u}$ and $\vec{\eta}_m \cdot \vec{v}$ are the logical bit and phase flip errors resulting when this Pauli operator is applied to an encoded state like \eqref{eq:encodedstabstate}.
In other words, the maximally-entangled state of $m$ physical qubits is the equal superposition
of a logical maximally-entangled state in all the possible encodings, which can be seen using the completeness of $\vec{\eta}_i$. Meanwhile, the
formulation of physical $X$ and $Z$ operators in terms of their logical versions follows from 
using the orthogonality of the $\vec{\eta}_i$ and $\vec{\xi}_j$.
Using these two facts we find that, after Bob's calculation of the relative syndrome $\vec{s}$, the tripartite state
can be expressed as (up to a local unitary acting only on Eve's systems)
\begin{equation}\label{eq:sigmapp}
\ket{\sigma''}_{ABE}=
  \sum_{\vec{u},\vec{v},\vec{f}} \sqrt{ p_{\vec{u},\vec{v}} q_{\vec{f}} }
 X^{ \vec{\xi}_m  \cdot (\vec{u}+\vec{f}) }_{B} Z^{ \vec{\eta}_m \cdot \vec{v} }_{B}
 \ket{\Phi^+}_{AB} \ket{\vec{f}}_{\mathbf{A'}}
 \ket{\vec{u}}_{E_1} Z^{\vec{f}}_{E_2}\ket{\vec{v}}_{E_2}
 \ket{ \vec{s} }_{\mathbf{B'}},
\end{equation}
where
$\vec{s}=( \vec{\xi}_1\cdot(\vec{u}+\vec{f}),\dots,\vec{\xi}_{m-1}\cdot(\vec{u}+\vec{f}) )$.
While the registers $A$ and $B$ in equation \eqref{eq:sigmap} have been $m$-qubit registers, here they contain only a single qubit each.
Alice missing $(m-1)$-qubits have been traced out since they contained only classical information about her absolute syndrome (accessible to all parties).
The rest of Bob's $m$-qubit register now contains classical information about the relative syndrome $\vec{s}$ and is labeled $\mathbf{B'}$.
We now restrict ourselves to the cat code, which is given by 
$(\vec{\xi}_i)_j=\delta_{1j}+\delta_{i+1,j}$ for $i=1\dots m-1$, $(\vec{\xi}_m)_j=\delta_{1j}$ and
$(\vec{\eta}_i)_j=\delta_{i+1,j}$ for $i=1\dots m-1$, $(\vec{\eta}_m)_j=1$ (see figure \ref{fig:cat}).
\begin{figure}[t]
\centering
\begin{pspicture}(-1.05,0.35)(4.375,-1.925)
\scalebox{0.7}{
\psframe[linecolor=lightgray,linestyle=dotted](-0.25,0.25)(2.75,-1.75) 
\pspolygon[linecolor=lightgray,linestyle=dashed](-0.5,0.5)(3,0.5)(3,-1.75)(6.25,-1.75)(6.25,-2.75)(-0.5,-2.75) 
\psframe[linecolor=lightgray,fillstyle=solid](0,0)(2.5,-1.5)
\rput[c](1.25,-0.75){
$\begin{array}{cccc}
 Z & Z & \id & \id \\
 Z & \id & Z & \id \\
 Z & \id & \id & Z
 \end{array}$}
\psframe[linecolor=lightgray,fillstyle=solid](0,-2)(2.5,-2.5)
\rput[c](1.25,-2.25){
$\begin{array}{cccc}
  Z & \id & \id & \id
 \end{array}$}
\psframe[linecolor=lightgray,fillstyle=solid](3.5,0)(6,-1.5)
\rput[c](4.75,-0.75){
$\begin{array}{cccc}
 \id & X & \id & \id \\
 \id & \id & X & \id \\
  \id & \id & \id & X
 \end{array}$}
\psframe[linecolor=lightgray,fillstyle=solid](3.5,-2)(6,-2.5)
\rput[c](4.75,-2.25){
$\begin{array}{cccc}
 X & X & X & X
 \end{array}$}
\rput[r](-0.05,-0.75){$m-1 \left\{ \makebox(0,0.9)[b]{}\right.$}
\rput[B](1.25,0.6){$m$}
\rput[B](1.25,0.1){\rotatebox[origin=c]{-90}{$\left\{ \makebox(0,1.35)[b]{}\right.$}}
}
\end{pspicture}
\vspace*{13pt}
\fcaption{\label{fig:cat}
Cat code encoding one qubit into $m=4$.
The operators on the left hand side are the $Z(\vec{\xi}_i) = Z^{\vec{\xi}_i}$ ($i=1\dots m$ from top to bottom),
those on the right hand side the $X(\vec{\eta}_i) = X^{\vec{\eta}_i}$.
The (generators of the) stabilizers are within the dotted line, the (generators of the) normalizers within the dashed one.}
\end{figure}
The name comes from the fact that 
$\alpha \ket{\theta(\vec{0},0)} + \beta \ket{\theta(\vec{0},1)} =
 \alpha \ket{00\dots0} + \beta \ket{11\dots1}$, a Schr\"odinger cat state when $\alpha=\beta=\frac{1}{\sqrt{2}}$.

Finally, Alice and Bob both measure their key bit.
Alice forgets about which bits she flipped by tracing out the $\mathbf{A'}$ register.
The correlations between Alice, Bob, and Eve are described by the following semiclassical state:
\begin{multline}
 \sigma_{XY\overline{E}} =
 \frac{1}{2} \sum_x [x]_A \otimes
 \sum_{\vec{u},\vec{f}} \sum_{\vec{v}_1,\vec{v}_2}
 \sqrt{ p_{\vec{u},\vec{v}_1} p_{\vec{u},\vec{v}_2} }  q_{\vec{f}} \,
 [x+\vec{\xi}_m \cdot(\vec{u}+\vec{f})]_B    \\
\otimes\, [ \vec{s} ]_{B'} \otimes  [\vec{u}]_{E_1} \otimes
 (Z^{\vec{\eta}_m})^x_{E_2} Z^{\vec{f}}_{E_2}
     \ketbra{\vec{v}_1}{\vec{v}_2}
 Z^{\vec{f}}_{E_2} (Z^{\vec{\eta}_m})^x_{E_2},
\end{multline}
where $[x]_B=\ket{x}\bra{x}_B$, etc.
Note that the state is diagonal in $E_1$ since the quantities $\vec{\xi}_i\cdot(\vec{u}+\vec{f})$ are all classical:
$i=1,\dots,m-1$ is already classical in \eqref{eq:sigmapp}, $i=m$ became classical after the key bit measurements by Alice and Bob.
The $\{ \vec{\xi}_i \}_{i=1,\dots m}$ span $\F^m$ thereby completely fixing the string $\vec{u}+\vec{f}$.

To calculate the quantum mutual information between Alice and Bob we trace out Eve and obtain
\begin{align}
 \sigma_{XY} &=
 \frac{1}{2} \sum_x [x]_A \otimes
 \sum_{\vec{u},\vec{f}} p_{\vec{u}} q_{\vec{f}} \,
 [x+\vec{\xi}_m \cdot(\vec{u}+\vec{f})]_B \otimes
 [  (\vec{\xi}_1 \cdot(\vec{u}+\vec{f}),\dots) ]_{B'} \nonumber\\
 &=
 \frac{1}{2} \sum_x [x]_A \otimes
 \sum_{\vec{u}} \tilde{p}_{\vec{u}}  \,
 [x+\vec{\xi}_m \cdot\vec{u}]_B \otimes
 [  (\vec{\xi}_1\cdot\vec{u},\vec{\xi}_2\cdot\vec{u},\dots) ]_{B'} \nonumber\\
 &=\frac{1}{2} \sum_x [x]_A \otimes \sum_{l_x, \vec{s}} \tilde{P}(l_x,\vec{s})
   [x+l_x]_B \otimes [ \vec{s} ]_{B'},
\end{align}
where $\tilde{p}=p(1-q)+(1-p)q$,
$\tilde{p}_{\vec{u}}$ is defined as $\tilde{p}_{\vec{u}}=\tilde{p}^u(1-\tilde{p})^{m-u}$, and
$\tilde{P}(l_x,\vec{s}) =
(\tilde{p}^s(1-\tilde{p})^{m-s})^{1-l_x}
(\tilde{p}^{m-s}(1-\tilde{p})^s)^{l_x}$.
In the last step we used $\vec{u} = l_x \vec{\eta}_m +\sum_i s_i \vec{\eta}_i$ to write the sum over $\vec{u}$ as a sum over $l_x$ and $\vec{s}$, where $l_x$ is the logical $X$ error, i.e.~$X$ error on the first qubit in the block.
This immediately yields 
\begin{equation}\label{eq:iab}
 I(X:Y) = 1 - \sum_{\vec{s}} \tilde{P}(\vec{s}) \bentropy{\tilde{P}(l_x\vert \vec{s})},
\end{equation}
using the binary entropy $\bentropy{x}=-x\log x-(1-x)\log(1-x)$.
Note that $I(X:Y)$ does not depend on the particular values $\{p_{uv}\}$ in $\sigma_{AB}$ (see \eqref{eq:sigma_ab}),
but only depends on the bit error rate $p=p_{10}+p_{11}$.
The form of the mutual information indicates the advantage provided by the syndrome.
If Alice did not send any information, Bob's state would be averaged over the possible syndromes, and
the mutual information would involve the entropy of the average of the $\tilde{P}(l_x\vert \vec{s})$ rather
than the average of the entropies. By concavity of entropy, the latter rate is larger.

To calculate the quantum mutual information between Alice and Eve, we trace out Bob's systems and obtain
\begin{align}
 \sigma_{X\overline{E}} &=  \frac{1}{2} \sum_x [x]_A \otimes \rho_{E_1E_2}^{(x)}, \\
 \rho_{E_1E_2}^{(x)} &=
 \sum_{\vec{u}} p_{\vec{u}} \, [\vec{u}]_{E_1} \otimes \rho_{E_2}^{(x),\vec{u}} , && \text{and} \label{eq:14}\\
 \rho_{E_2}^{(x),\vec{u}} &=
 (Z^{\vec{\eta}_m})^x
 \sum_{\vec{f}} q_{\vec{f}}
   Z^{\vec{f}}  \ketbra{ \Psi_{\vert\vec{u}} }{ \Psi_{\vert\vec{u}} }  Z^{\vec{f}}
 (Z^{\vec{\eta}_m})^x,
\end{align}
with
\begin{equation}
\ket{ \Psi_{\vert\vec{u}} } = \sum_{\vec{v}}\sqrt{ p_{\vec{v}\vert \vec{u}} } \ket{\vec{v}}.
\end{equation}
We proceed with the computation of the secure key rate for the BB84 and the 6-state protocol separately in the following two subsections.

\subsubsection{BB84}\label{sec:bb84}
To calculate the secure key rate we must find the minimum over all $\sigma_{AB}$ of the difference between the quantum mutual information between Alice and Bob and Alice and Eve.
Since $I(X:Y)$ does not depend on the particular structure of $\{p_{uv}\} = \{1-2p+t,p-t,t,p-t\} $, $t\in[0,p]$, in $\sigma_{AB}$, but only depends on the bit error rate $p=p_{10}+p_{11}$,
this corresponds to finding the maximum of $I(X:E)$.

Let us assume for a moment that this maximum is achieved for independent bit and phase errors, i.\,e.
we consider the state $\sigma_{AB}$ with $\{ p_{uv} \} = \{1-2p+t,p-t,t,p-t\}$ and $t=p^2$.
In this case $\ket{ \Psi_{\vert\vec{u}} }$ does not depend on $\vec{u}$, and we get
\begin{equation}
\rho_{E_2}^{(x),\vec{u}} =
 (Z^{\vec{\eta}_m})^x
 \rho_{pq}^{\otimes m}
 (Z^{\vec{\eta}_m})^x
\end{equation}
with $\rho_{pq} = (1-q)\ketbra{\varphi_+}{\varphi_+} + q \ketbra{\varphi_-}{\varphi_-}$ and
$\ket{\varphi_\pm} = \sqrt{1-p}\ket{0}\pm \sqrt{p}\ket{1}$.
Part $E_1$ and $E_2$ of the state $\rho^{(x)}_{E_1E_2}$ in \eqref{eq:14} are now completely decoupled.
As it was shown in \cite{SRS06}, the fact that $E_1$ is classical allows the corresponding state describing dependent errors to be reconstructed from this state:
After tracing out the $E_1$ part, we add an ancilla $[0]_{E_3}$, apply the isometry
$\sum_{\vec{u},\vec{v}}\sqrt{p_{\vec{u}\vert\vec{v}}} \ket{\vec{u}}_{E_3}\bra{0} \otimes [\vec{v}]_{E_2}$ and eventually dephase the ancilla.
Since quantum mutual information never increases under local operations, the maximum of $I(X:E)$ is indeed achieved for independent errors and we get
\begin{equation}\label{eq:iaebb84}
 I(X:E) = S\bigl( \frac{1}{2} \rho_{pq}^{\otimes m} + \frac{1}{2} (Z\rho_{pq}Z)^{\otimes m} \bigr)
 -m S\bigl( \rho_{pq} \bigr).
\end{equation}
Subtraction of \eqref{eq:iaebb84} from \eqref{eq:iab} gives the secure key rate for the BB84 protocol:
\begin{multline}\label{eq:ratebb84}
 r(m,p) = \max_q \frac{1}{m} \Bigl[ 1- \sum_{s=0}^{m-1} \binom{m-1}{s} \tilde{P}(s) \bentropy{\tilde{P}(l_x\vert s)} \\
- S\bigl( \frac{1}{2} \rho_{pq}^{\otimes m} + \frac{1}{2} (Z\rho_{pq}Z)^{\otimes m} \bigr) +m \bentropy{\frac{1}{2}(1+\sqrt{1-16p(1-p)q(1-q)})}  \Bigr].
\end{multline}
Omitting the maximization over $q$, the above formula gives the key rate $r_{m,q}(p)$ for some fixed values of $m$ and $q$ as a function of the bit error rate $p$.
By setting $r_{m,q}(p)$ equal to zero, we find $p_{max}(m,q)$, the maximum tolerable bit error rate for given $m$ and $q$.
For very high levels of added noise, i.\,e. for $q=\frac{1}{2}-\epsilon$, we find that
for all values of $m$, the key rate becomes zero at the bit error rate 
$p_{max}(m,q=\frac{1}{2}-\epsilon )=\BBPn$, but by adding less 
noise at higher values of $m$, secret keys can be generated for even larger bit error rates.

Without the use of the cat code (i.\,e. if we take $m=1$) the rate reduces to \cite{KGR05,KGR05b}
\begin{equation}\label{eq:ratebb84:m=1}
 r(p) = \max_q  \Bigl[ 1- \bentropy{\tilde{p}} - \bentropy{p} + \bentropy{\frac{1}{2}(1+\sqrt{1-16p(1-p)q(1-q)})}  \Bigr].
\end{equation}
With neither local randomization ($q=0$) nor use of the cat code ($m=1$) the key rate \eqref{eq:ratebb84:m=1} becomes even smaller \cite{SP00},
\begin{equation}\label{eq:ratebb84:m=1:q=0}
 r(p) =  1 - 2\bentropy{p},
\end{equation}
and secure key generation becomes impossible for bit error rates higher than $p_{max}=\BBP$.

\begin{figure}[t]
 \centering
 \includegraphics[scale=0.8]{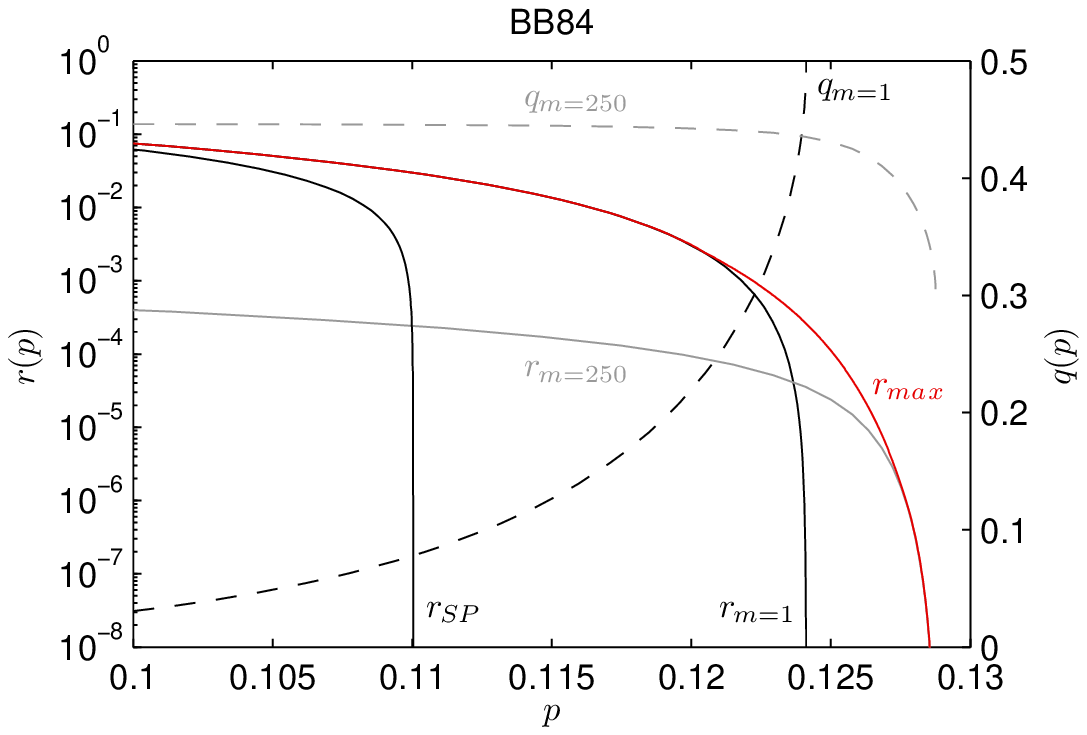}
 \vspace*{13pt}
 \fcaption{Secure key rate $r$ of BB84 for various types of preprocessing versus bit error rate $p$. No preprocessing corresponds to $r_{SP}$, noisy preprocessing to $r_{m=1}$, and the maximum over all block lengths $m\leq 250$ to $r_{max}$, shown in red.
For the rates achieved by the blocklengths $m=1$ and $m=250$, the corresponding rate of the added noise is shown on the right y axis.
\label{fig:bb84}}
\end{figure}
Figure \ref{fig:bb84} shows plots of the key rates given by \eqref{eq:ratebb84:m=1:q=0} and \eqref{eq:ratebb84:m=1} (black) and the maximum over the key rates given by \eqref{eq:ratebb84} (red) for values of $m$ up to $250$.
The increase of the maximal tolerable bit error rate with the block length $m$ is illustrated in figure \ref{fig:1}.
The highest value of $m$ for which we maximized the tolerable bit error rate as function of the added noise $q$ was $m=\BBPcm$ leading to $p_{max}(m=500,q=\BBPcq)=\BBPcr$.

By far the most difficult part in the numerical evaluation of \eqref{eq:ratebb84} is 
computing the von Neumann entropy, as it 
contains a sum of two $m$-fold tensor products of different one qubit density operators.
Such an expression can be more efficiently calculated by taking into account its block diagonal structure which follows from permutation invariance, as detailed in the Appendix.

\subsubsection{6-state}
Since the set $\Gamma$ only contains the single state $\{ p_{uv} \} = \{ 1-\frac{3}{2}p, \frac{p}{2}, \frac{p}{2}, \frac{p}{2}\}$, minimization over $\sigma_{AB}$ is unnecessary and the secure 
key rate is directly given by the difference of the quantum mutual informations between Alice and Bob \eqref{eq:iab} and Alice and Eve. Despite the simplicity of $\Gamma$, this calculation is more difficult than
BB84 due to the correlation between bit and phase errors. 
The corresponding conditional probabilities are given by
$p_{v=1\vert u=0} = \frac{p}{2(1-p)} = p'$, $p_{v=0\vert u=0} = 1-p'$
and $p_{v\vert u=1} = \frac{1}{2}$.
Therefore, denoting the number of ones in $\vec{u}$ as $u$ and by reordering the qubits in such a way that the first $u$ qubits are the ones with $u_i=1$,
we get
\begin{equation}
\ket{\Psi_{\vert\vec{u}}} = \sum_{\vec{v}}\sqrt{ p_{\vec{v}\vert \vec{u}} } \ket{\vec{v}}
          = \ket{+}^{\otimes u} \otimes \ket{\varphi_+'}^{\otimes m-u} = \ket{\Psi_{\vert u}}
\end{equation}
with $\ket{\pm}=\frac{1}{\sqrt{2}} ( \ket{0}\pm\ket{1} )$ and
$\ket{\varphi'_\pm} = \sqrt{p'}\ket{0} \pm \sqrt{1-p'}\ket{1}$, leading to
\begin{align}
 \rho_{E_2}^{(x),u} &=
 (Z^{\vec{\eta}_m})^x
 \sum_{\vec{f}} q_{\vec{f}}
   Z^{\vec{f}} [+]^{\otimes u} \otimes [\varphi'_+]^{\otimes m-u}  Z^{\vec{f}}
 (Z^{\vec{\eta}_m})^x \nonumber\\
 &= (Z^{\vec{\eta}_m})^x \sigma^{\otimes u} \otimes \gamma^{\otimes m-u} (Z^{\vec{\eta}_m})^x
\end{align}
with $\sigma=(1-q)[+]+q[-]$ and $\gamma=(1-q)[\varphi'_+]+q[\varphi'_-]$.
Reordering the state in 
this manner does not change the entropy, and so will not alter the rate.
Using these results the quantum mutual information between Alice and Eve can be expressed as
\begin{multline}\label{eq:iae6state}
 I(X:E) = \sum_{u=0}^m \binom{m}{u} p^u(1-p)^{m-u} \Bigl[
 S\bigl( \frac{1}{2}\sigma^{\otimes u}\otimes\gamma^{\otimes m-u} +
         \frac{1}{2}(Z\sigma Z)^{\otimes u}\otimes(Z\gamma Z)^{\otimes m-u} \bigr) \\
 -u \bentropy{q} - (m-u) \bentropy{\frac{1}{2}(1+\sqrt{1-16p'(1-p')q(1-q)}) }  \Bigr].
\end{multline}
Since $\sigma$ and $Z\sigma Z$ are diagonal in the same basis we are able to write the von Neumann entropy as
\begin{equation}
\sum_{k=0}^u \binom{u}{k} S\Bigl( \frac{(1-q)^kq^{u-k}}{2} \gamma^{\otimes m-u} +
     \frac{q^k(1-q)}{2}^{u-k}(Z\gamma Z)^{\otimes m-u}  \Bigr)
\end{equation}
which is of the same form as the von Neumann entropy in \eqref{eq:iaebb84}.
Therefore the same methods for evaluation can be applied; see the Appendix.

The secure key rate is given by subtracting \eqref{eq:iae6state} from \eqref{eq:iab},
\begin{equation}\label{eq:rate6state}
 r(m,p) = \max_q \frac{1}{m} \Bigl[ 1- \sum_{s=0}^{m-1} \binom{m-1}{s} \tilde{P}(s)
  \bentropy{\tilde{P}(l_x\vert s)} - I(X:E)  \Bigr].
\end{equation}
As it is the case for the BB84 protocol, the key rate becomes zero for all values of $m$ for $q\rightarrow \frac{1}{2}$ (this time at bit error rate $p_{max}(m,q=\frac{1}{2}-\epsilon)=\SSPn$), but again adding less noise at higher values of $m$ gives rise to secret keys for even higher bit error rates. 

Two special cases emerge from \eqref{eq:rate6state}, $m=1$:
\begin{equation}\label{eq:rate6state:m=1}
 r(p) = \max_q  \Bigl[ 1- \bentropy{\tilde{p}}
- \sum_u p_u \left( \bentropy{p_{v\vert u}} - \bentropy{\frac{1}{2}(1+\sqrt{1-16p_{1\vert u}(1-p_{1\vert u})q(1-q)}) } \right) \Bigr],
\end{equation}
and $q=0$, which leads to \cite{ShSm96,DiSS98,Lo01}:
\begin{equation}\label{eq:rate6state:q=0}
 r(p) = \frac{1}{m} \Bigl[ 1 - \sum_{\vec{s}} P(\vec{s}) \entropy{P(l^x,l^z\vert\vec{s}}  \Bigr]
\end{equation}
(note the appearance of the usual entropy, not the binary entropy), with
\begin{multline}
P(l^x,l^z,\vec{s}) = \frac{1}{2} \Bigl[
p^{l^x(m-2s)+s} (1-p)^{(1-l^x)(m-2s)+s} +  \\
 \delta_{0,l^x(m-2s)+s} \ (-1)^{l^z} (1-2p)^{(1-l^x)(m-2s)+s}
\Bigr]
\end{multline}
which attains the highest robustness for $m=5$ leading to $p_{max}(m=5,q=0)=\SSPd$ instead of $p_{max}(m=1,q=0)=\SSP$.
\footnote{Note that in \cite{ShSm96,DiSS98} the quantity under consideration was the capacity of the quantum depolarizing channel with $\{ p_{uv} \} = \{ 1-p, \frac{p}{3}, \frac{p}{3}, \frac{p}{3}\}$ therefore leading to $p_{max}=\frac{3}{2} \times \SSPd$}.

\begin{figure}[t]
 \centering
 \includegraphics[scale=0.8]{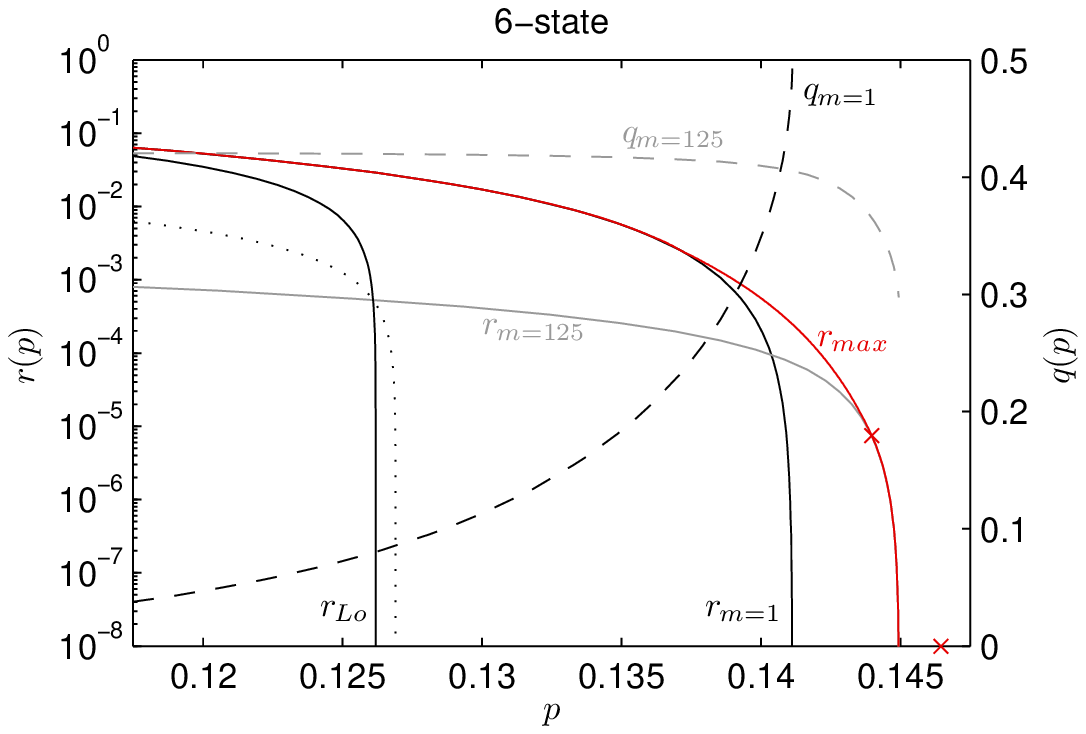}
 \vspace*{13pt}
 \fcaption{Secure key rate $r$ of the 6-state protocol for various types of preprocessing versus bit error rate $p$. No preprocessing corresponds to $r_{\rm Lo}$, noisy preprocessing to $r_{m=1}$, and the maximum achievable rate over all blocklengths $m\leq 125$, to $r_{max}$, shown in red. 
For the rates achieved by the blocklengths $m=1$ and $m=125$, the corresponding rate of the added noise is shown on the right y-axis.
The dotted rate with $p_{max}=\SSPd$ is due to Lo, corresponding to use of a repetition code of blocklength $m=5$ and no noisy preprocessing.
\label{fig:6state}}
\end{figure}

In figure \ref{fig:6state} we show the key rates in these special cases as well as the general case for optimal noise and blocklengths up to $m=125$.
Included are $q=0,m=1$ (black),
$q=0,m=5$ (dotted),
and $m=1$ for the optimal $q$ (black).
The maximum over the key rates given by \eqref{eq:rate6state} for values of $m$ up to $125$ is shown in red, along with the specific case of $m=125$.
The increase of the maximal tolerable bit error rate with the block length $m$ is illustrated in figure \ref{fig:1}.
The highest value of $m$ for which we maximized the tolerable bit error rate as function of the added noise $q$ was $m=\SSPcm$ leading to $p_{max}(m=\SSPcm,q=\SSPcq) = \SSPcr$.
Since the computation for larger blocksizes becomes rather slow, we extrapolated the value for the optimum noise leading to $q\approx \SSPeq$ for $m=\SSPem$.
By calculating the highest tolerable bit error for this value of noise we get the best lower bound $p_{max}(m=\SSPem,q=\SSPeq)=\SSPer$.

\section{Iterated Preprocessing}

By combining local randomization with the cat-code of 
size $m$, Alice and Bob gain an advantage over Eve and 
intuitively it seems this advantage might be even bigger by performing the procedure twice.
In this section we discuss such a twofold iterated protocol where Alice adds noise at a rate $q$ to $m_2$ blocks of size $m_1$ each, and then after measuring the syndromes of these blocks, adds further noise at another rate $Q$ to the $m_2$ 'key' bits of these blocks.
Then the syndrome of these $m_2$ bits is measured and the remainder of the protocol
proceeds as usual.
We restrict ourselves to the BB84 protocol for simplicity.
Using essentially the same argument as in section \ref{sec:bb84},
we find that we only need to consider independent bit and phase errors described by the state $\sigma_{AB}$ with $\{ p_{uv} \} = \{1-2p+t,p-t,t,p-t\}$ and $t=p^2$:
(i) $I(X:Y)$ depends only on the bit error rate $p=p_{10}+p_{11}$,
(ii) therefore we have to find the maximum of $I(X:E)$,
(iii) which is achieved for independent errors.
The proof of (iii) works as in section \ref{sec:bb84}, since, as we will see, $E_1$ of $\sigma_{X\overline{E}}$ is again classical.

\subsection{Rate Calculation}
We start with \eqref{eq:sigmapp}, now denoting $m$ as $m_1$.
Adding additional noise at rate $Q$ to the key bit the state is described as
\begin{multline}
\ket{\sigma'}_{ABE} =
 \sum_{\vec{f},\vec{u},\vec{v}}
 \sum_F
 \sqrt{ p_{\vec{u},\vec{v}} q_{\vec{f}}\, Q_F}
 X^{ \vec{\xi}_{m_1}\cdot(\vec{u}+\vec{f}) + F }_{B} Z^{ \vec{\eta}_{m_1}\cdot\vec{v} }_{B}
 \ket{\Phi^+}_{AB} \\
 \otimes  \ket{\vec{u}}_{E_1} 
 (Z^{\vec{\eta}_{m_1}})^F Z^{\vec{f}} \ket{\vec{v}}_{E_2} 
 \ket{ \vec{s}\, }_{\mathbf{B}} \ket{\vec{f}\,}_{\mathbf{A'}} \ket{F}_{\mathbf{A''}}
\end{multline}
with classical registers $\mathbf{B}$, $\mathbf{A'}$ and $\mathbf{A''}$
and the same relationship 
$\vec{s}=( \vec{\xi}_1\cdot(\vec{u}+\vec{f}),\dots,\vec{\xi}_{m_1-1}\cdot(\vec{u}+\vec{f}) )$ as before.
Now we consider the $m_2$-fold tensor product $\ket{\sigma}_{ABE}^{'\otimes m_2}$ and define the abbreviations
$\vec{U} = ( \vec{\xi}_{m_1}\cdot(\vec{u}_1+\vec{f}_1), \dots, \vec{\xi}_{m_1}\cdot(\vec{u}_{m_2}+\vec{f}_{m_2}) )$ and
$\vec{V} = ( \vec{\eta}_{m_1}\cdot\vec{v}_1, \dots, \vec{\eta}_{m_1}\cdot\vec{v}_{m_2} )$.
Again Alice and Bob both measure their stabilizers, Alice sends her result to Bob, who calculates the relative syndrome
$\vec{S} = ( \vec{\xi}_1\cdot(\vec{U}+\vec{F}), \dots, \vec{\xi}_{m_2-1}\cdot(\vec{U}+\vec{F}) )$. Both then measure their key bit.
The tripartite semiclassical state describing the correlations is now given by
\begin{multline}
\sigma_{XY\overline{E}} = \frac{1}{2} 
 \sum_{\vec{F}} Q_{\vec{F}}
 \sum_{\vec{f}_1,\dots,\vec{f}_{m_2}} q_{\vec{f}_1}\dots q_{\vec{f}_{m_2}}
 \sum_{\vec{u}_1,\dots,\vec{u}_{m_2}}
 p_{\vec{u}_1}\dots p_{\vec{u}_{m_2}} \\
 \times\, \sum_x [x]_A \otimes [x+L_x]_B \otimes [\vec{s}_1,\dots,\vec{s}_{m_2},\vec{S}]_{\mathbf{B}} \otimes   [\vec{u}_1,\dots,\vec{u}_{m_2}]_{E_1} \\
\otimes \,
 ( Z^{\otimes m_1m_2} )^x
  \bigotimes_{i=1}^{m_2}
  \bigl(
   (Z^{\otimes m_1})^{F_i} Z^{\vec{f}_i} \ketbra{\Psi}{\Psi} Z^{\vec{f}_i} (Z^{\otimes m_1})^{F_i}
  \bigr)
 ( Z^{\otimes m_1m_2} )^x,
\end{multline}
where $\ket{ \Psi } = \sum_{\vec{v}} \sqrt{ p_{\vec{v}} } \ket{\vec{v}}$
and $\vec{s}_i=(\vec{\xi}_1\cdot(\vec{u}_i+\vec{f}_i),\dots)$,
 $\vec{S}=(\vec{\xi}_1\cdot(\vec{U}+\vec{F}),\dots)$, and
$L_x=\vec{\xi}_{m_2}\cdot(\vec{U}+\vec{F})$.

To calculate the quantum mutual information between Alice and Bob we trace out Eve's systems and obtain
\begin{equation}
\sigma_{XY} = \frac{1}{2} 
 \sum_{\vec{F}} Q_{\vec{F}}
 \sum_{\vec{u}_1 \dots \vec{u}_{m_2}}
 \tilde{p}_{\vec{u}_1}\dots \tilde{p}_{\vec{u}_{m_2}}\sum_x\, [x]_A \otimes
 [x+L_x]_B \otimes [\vec{s}_1 \dots \vec{s}_{m_2},\vec{S}]_B,
\end{equation}
using $\tilde{p}=p(1-q)+(1-p)q$. Since Alice's additional noise $\vec{f}$ is now combined with Eve's noise
$\vec{u}$, $\vec{f}$ no longer appears in the the syndromes $\vec{s}_i$ and $\vec{S}$:  
$\vec{s}_i=(\vec{\xi}_1\cdot\vec{u}_i,\dots)$, $\vec{S}=(\vec{\xi}_1\cdot(\vec{U}'+\vec{F}),\dots)$.  Additionally, $L_x$ is now $L_x=\vec{\xi}_{m_2}\cdot(\vec{U}'+\vec{F}),$
with $\vec{U}' = ( \vec{\xi}_{m_1}\cdot\vec{u}_1, \dots, \vec{\xi}_{m_1}\cdot\vec{u}_{m_2} )$.
The mutual information can therefore be written as
\begin{equation}\label{eq:iab:iter}
 I(X:Y) = 1 -
 \sum_{\vec{s}_1 \dots \vec{s}_{m_2},\vec{S}}
 \tilde{P}'(\vec{s}_1 \dots \vec{s}_{m_2},\vec{S})
 \bentropy{\tilde{P}'(L_x\vert \vec{s}_1 \dots \vec{s}_{m_2},\vec{S})},
\end{equation}
where the probability distribution $\tilde{P}'$ only depends on the number of ones 
in each of the syndromes $\vec{s}_i$ and $\vec{S}$
(we assume that the zeros and ones in $\vec{S}$ are ordered such that the syndromes $\vec{s}_i$, $i\in\{1,\dots,m_2-S\}$, correspond to $S_i=0$):
\begin{multline}\label{eq:itprob}
 \tilde{P}'( L_x=0, s_1\dots s_{m_2}, S) =
 \prod_{i=1}^{m_2-S}[ (1-\tilde{p})^{m_1-s_i} \tilde{p}^{s_i} (1-Q)
                     +(1-\tilde{p})^{s_i} \tilde{p}^{m_1-s_i} Q ] \times \\
 \prod_{i=m_2-S+1}^{m_2}[ (1-\tilde{p})^{s_i} \tilde{p}^{m_1-s_i} (1-Q)
                         +(1-\tilde{p})^{m_1-s_i} \tilde{p}^{s_i} Q ]  .
\end{multline}
In addition we see from \eqref{eq:itprob} that for a given value of $S$ only the frequency distribution of the
$s_i$, $i\in\{1,\dots,m_2-S\}$, and the $s_j$, $j\in\{m_2-S+1,\dots,m_2\}$, matters.
This fact can be used to speed up the calculation of the sum over the syndromes in \eqref{eq:iab:iter}.

Since $\rho_{E_2}^{(x),\vec{u}} =( Z^{\otimes m_1m_2} )^x
  \bigl[
    (1-Q) \rho_{pq}^{\otimes m_1} + Q (Z\rho_{pq}Z)^{\otimes m_1}
  \bigr]^{\otimes m_2}
 ( Z^{\otimes m_1m_2} )^x$,
the mutual information between Alice and Eve can be seen to be
\begin{multline}\label{IAE:BB84:it}
 I(X:E) =  S\Bigl(
  \frac{1}{2}\bigl[ (1-Q)\rho_{pq}^{\otimes m_1} + Q(Z\rho_{pq}Z)^{\otimes m_1} \bigr]^{\otimes m_2}
+ \frac{1}{2}\bigl[ Q\rho_{pq}^{\otimes m_1} + (1-Q)(Z\rho_{pq}Z)^{\otimes m_1} \bigr]^{\otimes m_2}
  \Bigr) \\
 -m_2 S\bigl( (1-Q)\rho_{pq}^{\otimes m_1} + Q(Z\rho_{pq}Z)^{\otimes m_1} \bigr)
\end{multline}
Once more the secure key rate is given by the difference of these mutual informations,
\begin{equation}\label{eq:ratebb84:it}
 r(m_1,m_2,p)= \max_{q,Q} \frac{1}{m_1m_2} \bigl(  I(X:Y)-I(X:E) \bigr).
\end{equation}

Again the hardest part in the numerical evaluation of \eqref{eq:ratebb84:it} comes from the von Neumann entropies.
One contains a sum of two $m_2$-fold tensor products of different density operators, but this time these density operators are $m_1$-qubit density operators.
Such an expression can also be calculated more efficiently by taking into account its permutation invariance.
For more details see the Appendix.

\begin{figure}[t]
 \centering
 \includegraphics[scale=0.8]{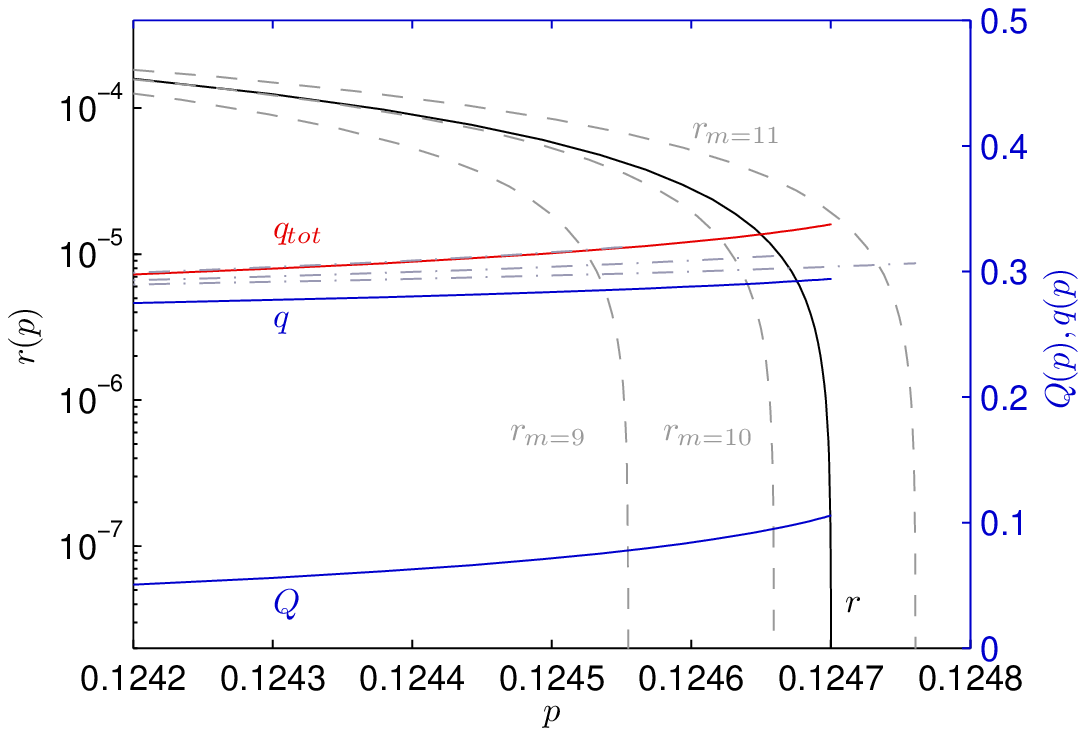}
 \vspace*{13pt}
 \fcaption{
Secure key rate $r$ of BB84 with iterated preprocessing of size $m_1\times m_2=3\times 3$ versus bit error rate $p$.
The right y-axis shows the corresponding values of added noise in the first ($q$) and second iteration ($Q$) as well as values of the total amount of added noise ($q_{tot}=q(1-Q)+(1-q)Q$, red). 
For comparison, the rates of the non-iterated protocol are shown for blocksizes $m\in\{9,10,11\}$ (dashed lines). The corresponding values of added noise for these cases are also shown (dash-dot lines).
\label{fig:bb84:it}}
\end{figure}

We compare the resulting key rate of the $m_1\times m_2=3\times 3$ iterated code with the key rates of the non-iterated codes of blocksizes $m\in\{9,10,11\}$ in figure \ref{fig:bb84:it}.

\begin{figure}[t]
\centering
\begin{pspicture}(-1.05,0.35)(11.375,-4.725)
\scalebox{0.7}{
\psframe[linecolor=lightgray,fillstyle=solid](0,0)(7.5,-5.5)  
\psframe[linecolor=lightgray,fillstyle=solid](0,-6)(7.5,-6.5) 
\psframe[linecolor=lightgray,fillstyle=solid](8.5,0)(16,-5.5) 
\psframe[linecolor=lightgray,fillstyle=solid](8.5,-6)(16,-6.5)

\psframe[linecolor=lightgray,linestyle=dotted](-0.25,0.25)(7.75,-5.75) 
\pspolygon[linecolor=lightgray,linestyle=dashed](-0.5,0.5)(8,0.5)(8,-5.75)(16.25,-5.75)(16.25,-6.75)(-0.5,-6.75) 
\psframe[linecolor=lightgray,fillstyle=solid](0,0)(2.5,-1.5)
\rput[c](1.25,-0.75){
$\begin{array}{cccc}
 Z & Z & \id & \id \\
 Z & \id & Z & \id \\
 Z & \id & \id & Z
 \end{array}$}
\psframe[linecolor=lightgray,fillstyle=solid](2.5,-1.5)(5,-3)
\rput[c](3.75,-2.25){
$\begin{array}{cccc}
 Z & Z & \id & \id \\
 Z & \id & Z & \id \\
 Z & \id & \id & Z
 \end{array}$}
\psframe[linecolor=lightgray,fillstyle=solid](5,-3)(7.5,-4.5)
\rput[c](6.25,-3.75){
$\begin{array}{cccc}
 Z & Z & \id & \id \\
 Z & \id & Z & \id \\
 Z & \id & \id & Z
 \end{array}$}
\psframe[linecolor=lightgray,fillstyle=solid](0,-4.5)(2.5,-5.5)
\rput[c](1.25,-5){
$\begin{array}{cccc}
 Z & \id & \id & \id \\
 Z & \id & \id & \id
 \end{array}$}
\psframe[linecolor=lightgray,fillstyle=solid](2.5,-4.5)(5,-5.5)
\rput[c](3.75,-5){
$\begin{array}{cccc}
 Z & \id & \id & \id \\
 \id & \id & \id &\id
 \end{array}$}
\psframe[linecolor=lightgray,fillstyle=solid](5,-4.5)(7.5,-5.5)
\rput[c](6.25,-5){
$\begin{array}{cccc}
 \id & \id & \id &\id \\
 Z & \id & \id & \id
 \end{array}$}

\psframe[linecolor=lightgray,fillstyle=solid](0,-6)(2.5,-6.5)
\rput[c](1.25,-6.25){
$\begin{array}{cccc}
  Z & \id & \id & \id
 \end{array}$}
\psframe[linecolor=lightgray,fillstyle=solid](2.5,-6)(5,-6.5)
\rput[c](3.75,-6.25){
$\begin{array}{cccc}
 \id & \id & \id & \id
 \end{array}$}
\psframe[linecolor=lightgray,fillstyle=solid](5,-6)(7.5,-6.5)
\rput[c](6.25,-6.25){
$\begin{array}{cccc}
 \id & \id & \id & \id
 \end{array}$}

\psframe[linecolor=lightgray,fillstyle=solid](8.5,0)(11,-1.5)
\rput[c](9.75,-0.75){
$\begin{array}{cccc}
 \id & X & \id & \id \\
 \id & \id & X & \id \\
  \id & \id & \id & X
 \end{array}$}
\psframe[linecolor=lightgray,fillstyle=solid](11,-1.5)(13.5,-3)
\rput[c](12.25,-2.25){
$\begin{array}{cccc}
 \id & X & \id & \id \\
 \id & \id & X & \id \\
  \id & \id & \id & X
 \end{array}$}
\psframe[linecolor=lightgray,fillstyle=solid](13.5,-3)(16,-4.5)
\rput[c](14.75,-3.75){
$\begin{array}{cccc}
 \id & X & \id & \id \\
 \id & \id & X & \id \\
  \id & \id & \id & X
 \end{array}$}
\psframe[linecolor=lightgray,fillstyle=solid](8.5,-4.5)(11,-5.5)
\rput[c](9.75,-5){
$\begin{array}{cccc}
 \id & \id & \id & \id \\
 \id & \id & \id & \id
 \end{array}$}
\psframe[linecolor=lightgray,fillstyle=solid](11,-4.5)(13.5,-5.5)
\rput[c](12.25,-5){
$\begin{array}{cccc}
 X & X & X & X \\
 \id & \id & \id & \id
 \end{array}$}
\psframe[linecolor=lightgray,fillstyle=solid](13.5,-4.5)(16,-5.5)
\rput[c](14.75,-5){
$\begin{array}{cccc}
 \id & \id & \id & \id \\
 X & X & X & X
 \end{array}$}

\psframe[linecolor=lightgray,fillstyle=solid](8.5,-6)(11,-6.5)
\rput[c](9.75,-6.25){
$\begin{array}{cccc}
 X & X & X & X
 \end{array}$}
\psframe[linecolor=lightgray,fillstyle=solid](11,-6)(13.5,-6.5)
\rput[c](12.25,-6.25){
$\begin{array}{cccc}
 X & X & X & X
 \end{array}$}
\psframe[linecolor=lightgray,fillstyle=solid](13.5,-6)(16,-6.5)
\rput[c](14.75,-6.25){
$\begin{array}{cccc}
 X & X & X & X
 \end{array}$}

\rput[r](-0.05,-0.75){$m_1-1 \left\{ \makebox(0,0.9)[b]{}\right.$}
\rput[r](-0.05,-5){$m_2-1 \left\{ \makebox(0,0.6)[b]{}\right.$}
\rput[B](1.25,0.6){$m_1$}
\rput[B](1.25,0.1){\rotatebox[origin=c]{-90}{$\left\{ \makebox(0,1.35)[b]{}\right.$}}
}
\end{pspicture}
\vspace*{13pt}
\fcaption{\label{fig:concat}
The concatenated code of size $m_1=4$ and $m_2=3$ encoding one qubit into $n=m_1\times m_2$.
The operators on the left hand side are the $Z(\vec{\xi}_i) = Z^{\vec{\xi}_i}$,
those on the right hand side the $X(\vec{\eta}_i) = X^{\vec{\eta}_i}$.
The stabilizers are within the dotted line, the normalizers within the dashed one.}
\end{figure}

\section{Conclusions}
Although sending quantum information and establishing a secret key are not equivalent uses of a quantum channel, 
their similarities are enough to ensure a fruitful exchange of techniques between the two problems.  
Here we have extended the application of structured codes to the problem of QKD given in~\cite{SRS06} 
to the 6-state protocol, showing how preprocessing based on 
both structured codes and local randomization leads to higher tolerable error rates. Having one more unbiased
basis implies correlation between bit and phase errors of the estimated quantum state, leading to increased
robustness of the protocol. Indeed, it seems likely 
that for large blocklength ($m\approx 500$) the threshold of the 6-state protocol 
exceeds the lowest known upper bound on the threshold for the BB84 protocol ($\BBPub$).
This method of preprocessing does not close the gap between BB84 upper and lower bounds, however,  
as the available data suggests that the BB84 threshold never exceeds roughly 13\% even for asymptotically-long blocks.
For the 6-state protocol, more data is needed to draw even a partial conclusion on the asymptotic threshold. 

Additionally, we have shown that further performance gains are possible when iterating the preprocessing step. 
Although our results are confined to two rounds with small block sizes, we can nevertheless already observe 
some general properties. The entire rate curve shifts to higher values, so this type of processing might
be useful in increasing the efficiency of protocols running over noisy channels. Intriguingly, the total amount of 
noise added to the sifted key bits is essentially the same as in the case of one round, showing that the 
improvement comes from making better use of the same amount of noise. More sophisticated 
representation-theoretic methods, in particular a Clebsch-Gordon decomposition of the states input to the second preprocessing round, should make analysis of more rounds and larger blocksizes tractable.

\nonumsection{Acknowledgements}
\noindent
We thank Gernot Alber and Graeme Smith for helpful discussions.
Financial support by the EC within the IP SECOQC is acknowledged. JMR acknowledges support from the
Alexander von Humboldt Foundation.

\nonumsection{References}
\noindent
\bibliography{6state}

\begin{thebibliography}{10}

\bibitem{BB84}
Charles~H. Bennett and Gilles Brassard.
\newblock Quantum cryptography: Public key distribution and coin tossing.
\newblock In {\em IEEE International Conference on Computers Systems and Signal
  Processing}, pages 175--179, Bangalore, India, 1984.

\bibitem{Bru98}
Dagmar Bru\ss.
\newblock Optimal eavesdropping in quantum cryptography with six states.
\newblock {\em Physical Review Letters}, 81:3018, October 1998.

\bibitem{SP00}
Peter~W. Shor and John Preskill.
\newblock Simple proof of security of the {BB}84 quantum key distribution
  protocol.
\newblock {\em Physical Review Letters}, 85:441, July 2000.

\bibitem{Lo01}
Hoi-Kwong Lo.
\newblock Proof of unconditional security of six-state quantum key distribution
  scheme.
\newblock {\em Quantum Information and Computation}, 1:81--94, August 2001.

\bibitem{devetak_relating_2004}
I.~Devetak and A.~Winter.
\newblock Relating quantum privacy and quantum coherence: An operational
  approach.
\newblock {\em Physical Review Letters}, 93:080501--4, 2004.

\bibitem{devetak_private_2005}
I.~Devetak.
\newblock The private classical capacity and quantum capacity of a quantum
  channel.
\newblock {\em IEEE Transactions on Information Theory}, 51:44-- 55, 2005.

\bibitem{devetak_distillation_2005}
Igor Devetak and Andreas Winter.
\newblock Distillation of secret key and entanglement from quantum states.
\newblock {\em Proceedings of the Royal Society A}, 461:207--235, 2005.

\bibitem{ShSm96}
Peter~W Shor and John~A Smolin.
\newblock Quantum error-correcting codes need not completely reveal the error
  syndrome.
\newblock {\em quant-ph/9604006v2}, April 1996.

\bibitem{DiSS98}
David~P. DiVincenzo, Peter~W. Shor, and John~A. Smolin.
\newblock Quantum-channel capacity of very noisy channels.
\newblock {\em Physical Review A}, 57:830, February 1998.

\bibitem{SmSm07}
Graeme Smith and John~A. Smolin.
\newblock Degenerate quantum codes for pauli channels.
\newblock {\em Physical Review Letters}, 98:030501--4, 2007.

\bibitem{KGR05}
B.~Kraus, N.~Gisin, and R.~Renner.
\newblock Lower and upper bounds on the secret-key rate for quantum key
  distribution protocols using one-way classical communication.
\newblock {\em Physical Review Letters}, 95:080501--4, 2005.

\bibitem{KGR05b}
Renato Renner, Nicolas Gisin, and Barbara Kraus.
\newblock Information-theoretic security proof for quantum-key-distribution
  protocols.
\newblock {\em Physical Review A}, 72:012332--17, July 2005.

\bibitem{horodecki_secure_2005}
Karol Horodecki, Michal Horodecki, Pawel Horodecki, and Jonathan Oppenheim.
\newblock Secure key from bound entanglement.
\newblock {\em Physical Review Letters}, 94:160502--4, April 2005.

\bibitem{RenesSmith07}
Joseph~M. Renes and Graeme Smith.
\newblock Noisy processing and distillation of private quantum states.
\newblock {\em Physical Review Letters}, 98:020502, 2007.

\bibitem{SRS06}
Graeme Smith, Joseph~M Renes, and John~A Smolin.
\newblock Better codes for {BB}84 with one-way post-processing.
\newblock {\em quant-ph/0607018}, July 2006.

\bibitem{FGGNP97}
Christopher~A. Fuchs, Nicolas Gisin, Robert~B. Griffiths, Chi-Sheng Niu, and
  Asher Peres.
\newblock Optimal eavesdropping in quantum cryptography. i. information bound
  and optimal strategy.
\newblock {\em Physical Review A}, 56:1163, 1997.

\bibitem{moroder_one-way_2006}
Tobias Moroder, Marcos Curty, and Norbert L{\"u}tkenhaus.
\newblock One-way quantum key distribution: Simple upper bound on the secret
  key rate.
\newblock {\em quant-ph/0603270}, March 2006.

\bibitem{phdrenner}
Renato Renner.
\newblock {\em Security of Quantum Key Distribution}.
\newblock PhD thesis, ETH Z{\"u}rich, 2006.

\bibitem{Ha03Fi}
Mitsuru Hamada.
\newblock Notes on the fidelity of symplectic quantum error-correcting codes.
\newblock {\em International Journal of Quantum Information}, 1:443--463,
  December 2003.

\bibitem{CPW02}
Jin-Quan Chen, Jialun Ping, and Fan Wang.
\newblock {\em Group Representation Theory for Physicists}.
\newblock World Scientific Publishing Company, 2nd edition, September 2002.

\end{thebibliography}

\appendix
\noindent
\section*{Efficient computation of the key rates}
\noindent
To evaluate the secure key rate of the BB84 and the 6-state quantum key distribution protocols involving our preprocessing protocol,
we need to compute the von Neumann entropy of a linear combination of $m$-fold tensor products of different one-qubit density matrices,
$S( \alpha\rho^{\otimes m} + \beta\sigma^{\otimes m} )$.
In the case of the iterated preprocessing protocol this expression becomes a sum over $m_2$-fold tensor products of some qudit density matrices of dimension $d=2^{m_1}$.
In this appendix we discuss how such expressions can be evaluated efficiently by using the Schur transform,
following the presentation in~\cite{CPW02}.

Considering $m$ qudits of dimension $d$, the Schur transform is a unitary transformation relating the standard computational basis
$\{ \ket{i_1,\dots i_m} \}$, $i_j=0,1,\dots,d-1$,
to a basis associated with the representation theory of the symmetric and general linear groups,
\begin{equation}\label{eq:schur}
\ket{\lambda}\ket{p_\lambda}\ket{q_\lambda} = \sum_{i_1\dots i_m}
  [U_{Sch}]^{\lambda p_\lambda q_\lambda}_{i_1\dots i_m} \ket{i_1, \dots , i_m}.
\end{equation}
The new basis $\{ \ket{\lambda}\ket{p_\lambda}\ket{q_\lambda} \}$ is labeled by a Young diagram $\lambda$ denoting the irreducible representations of both $\mathcal{S}_m$ and $\mathsf{GL}_d$,
a Young tableau $p_\lambda$ labeling the basis vectors spanning the representation spaces of $\mathcal{S}_m$,
and a Weyl tableau $q_\lambda$ labeling the the basis vectors spanning the representation spaces of $\mathsf{GL}_d$.
Elements $\sigma\in\mathsf{GL}_d$ and $s\in\mathcal{S}_m$ transform the basis states according to
\begin{align}
 \sigma^{\otimes m} \ket{\lambda}\ket{p_\lambda}\ket{q_\lambda} &=
      \ket{\lambda}\ket{p_\lambda} \, ( \sigma_\lambda \ket{q_\lambda} ),  \\
 s \, \ket{\lambda}\ket{p_\lambda}\ket{q_\lambda} &=
      \ket{\lambda} \, (s_\lambda \ket{p_\lambda}) \, \ket{q_\lambda},
\end{align}
whereas $\sigma_\lambda$ denotes $\sigma^{\otimes m}$ in the irrep $\lambda$ and
$s_\lambda$ denotes $s$ in the irrep $\lambda$.

It then follows that states like $\rho^{\otimes m}$ are block-diagonal, with 
blocks labeled by a Young diagram and tableau $(\lambda, p_\lambda)$.
Blocks with the same Young diagram are identical, and 
the number of Young tableaux $N_Y(\lambda)$ specifies degeneracy.
Meanwhile, the dimension of the blocks is given by the corresponding number of 
Weyl tableaux $N_W(\lambda)$.

This structure is clearly helpful for the calculation of expressions like
$S( \alpha\rho^{\otimes m} + \beta\sigma^{\otimes m} )$.
The entropy of the state reduces to the sum of entropies of the blocks, which itself factors into the 
entropy for a block $(\lambda ,p_{\lambda})$ with a fixed Young tableau $p_{\lambda}$, say the first one, times the corresponding degeneracy $N_Y(\lambda)$.

For $d>2$ we explicitly calculate both the $\rho_\lambda$ and the $\sigma_\lambda$ blocks using \eqref{eq:schur}, e.\,g.
\begin{multline*}
(\rho_\lambda)_{Qq} =
\bra{Q_\lambda}\bra{p_\lambda}\bra{\lambda} \rho^{\otimes m} \ket{\lambda}\ket{p_\lambda}\ket{q_\lambda}= \\
\sum_{j_1\dots j_m} \sum_{i_1\dots i_m}
 [U^\ast_{Sch}]^{\lambda p_\lambda Q_\lambda}_{j_1\dots j_m}
 [U_{Sch}]^{\lambda p_\lambda q_\lambda}_{i_1\dots i_m}
 \bra{j_1, \dots , j_m} \rho^{\otimes m} \ket{i_1, \dots , i_m}.
\end{multline*}
The Schur transform itself can be obtained using e.\,g. the eigenfunction method \cite{CPW02}.
For the iterated protocol it may be possible to further streamline the calculation by taking into
account the fact that the qudit inputs to the second round are themselves block-diagonal. 
Computations for much larger blocksizes $m_1\times m_2$ may then become feasible.

For $d=2$ things are simpler, as Young diagrams $\lambda$ consist of at most two rows 
and can be labeled by an index $j$, where $2j$ is the number of columns consisting of one row only ($j=0\dots \frac{m}{2}$ if $m$ is even, $j=\frac{1}{2}\dots \frac{m}{2}$ if $m$ is odd).
\begin{center}
\begin{pspicture}(-0.4,0)(4.8,1.44)
\scalebox{0.8}{
\rput[c](0.,1.5){$j$}
\rput[c](0.,1.0625){$\updownarrow$}
\rput[B](0.,0.52){$\lambda$}\rput[B](0.31,0.52){$=$}
\rput[t](1,1){\yng(2,2)}
\rput[c](1.8,0.6){$\dots$}
\rput[t](3,1){\yng(4,2)}
\rput[c](4.2,0.8){$\dots$}
\rput[t](5,1){\yng(2)}
\rput[c](1.8,1.5){$\frac{m}{2}-j$}
\rput[B](1.8,1.1){\rotatebox[origin=c]{-90}{$\left\{ \makebox(0,1.3)[b]{}\right.$}}
\rput[c](4.2,1.5){$2j$}
\rput[B](4.2,1.1){\rotatebox[origin=c]{-90}{$\left\{ \makebox(0,1.3)[b]{}\right.$}}  }
\end{pspicture}
\end{center}
The dimension of an irreducible representation with label $j$ is $N_W(j)=2j+1$ and the Weyl tableaux are now labeled $k=-j,\dots,j$, whereas $j+k$ denotes the number of ones in the first row of the Weyl tableaux:
\begin{center}
\begin{pspicture}(-0.4,0)(4.8,1.44)
\scalebox{0.8}{
\rput[t](1,1){\young(00,11)}
\rput[c](1.8,0.6){$\dots$}
\rput[t](3,1){\young(0000,11)}
\rput[c](4.2,0.8){$\dots$}
\rput[t](5,1){\young(11)}
\rput[c](3.625,0.1){$j-k$}
\rput[B](3.625,0.3){\rotatebox[origin=c]{90}{$\left\{ \makebox(0,0.67)[b]{}\right.$}}
\rput[c](4.825,0.1){$j+k$}
\rput[B](4.825,0.3){\rotatebox[origin=c]{90}{$\left\{ \makebox(0,0.67)[b]{}\right.$}}
\rput[c](1.8,1.5){$\frac{m}{2}-j$}
\rput[B](1.8,1.1){\rotatebox[origin=c]{-90}{$\left\{ \makebox(0,1.3)[b]{}\right.$}}
\rput[c](4.2,1.5){$2j$}
\rput[B](4.2,1.1){\rotatebox[origin=c]{-90}{$\left\{ \makebox(0,1.3)[b]{}\right.$}}  }
\end{pspicture}
\end{center}
The corresponding degeneracy follows from Robinson's hook length formula, which in this case yields
\begin{equation}
N_Y(j) = \binom{m}{m/2-j} \frac{2j+1}{m/2+j+1}.
\end{equation}
We diagonalize $\rho$ and $\sigma$, denoting the two eigenvalues of $\rho$ [$\sigma$] as $\rho_1$ and $\rho_2$  [$\sigma_1$ and $\sigma_2$],
\begin{align*}
 \rho &= U_\rho \varrho U^\dagger_\rho, &  \varrho&=\mathsf{diag}\{ \rho_1,\rho_2 \}, \\
 \sigma &= U_\sigma \varsigma U^\dagger_\sigma, & \varsigma&=\mathsf{diag}\{ \sigma_1,\sigma_2 \},
\end{align*}
and note that diagonal density operators like $\varrho$ and $\varsigma$ can easily be expressed in the $j$-th representation, since they are diagonal in all these representations, too.
The action of $\varrho^{\otimes m}$ on basis states \eqref{eq:schur} of the Schur basis becomes simply a multiplication by powers of the two eigenvalues because of the symmetry properties of these basis states:
Each basis state of the Schur basis labeled by a certain Young diagram $j$ and Weyl tableaux $k$ consists of a superposition of computational basis states which are permutations of
$\ket{01}^{\otimes(m/2-j)}\ket{0}^{\otimes(j-k)}\ket{1}^{\otimes(j+k)}$
independently of the Young tableaux (specifying degeneracy).
We obtain
\begin{equation}
\varrho_j = \mathsf{diag} \{
  \rho_1^{j-k}  \rho_2^{j+k}  (\rho_1 \rho_2)^{m/2-j}    \}_{k=-j\dots j} ,
\end{equation}
and an analogous expression for $\varsigma_j$.
To get the non-diagonal original block $\rho_j$ [$\sigma_j$] we have to apply the unitary $U_\rho$ [$U_\sigma$] in the irrep $j$ which is given by the Wigner rotation matrices $D_j( U_\rho )$ [$D_j( U_\sigma )$].
In our case these Wigner matrices are given simply by matrix exponentiation, $D_j( U_\rho ) = \exp( - \frac{\theta(\rho)}{2} (J_+-J_-) )$, where $\theta(\rho)$ denotes a phase calculated from $\rho$, and $J_\pm$ denotes the usual angular momentum ladder operators,
$J_\pm \ket{jm} =\sqrt{j(j+1)+ m(m\pm 1)}\ket{jm\pm 1}$.
In this way it becomes feasible to calculate expressions like $S( \alpha \sigma^{\otimes m} + \beta \rho^{\otimes m} )$ for values of $m$ up to several hundreds.

\end{document}